\documentclass[apj,numberedappendix]{emulateapj}
\usepackage{epstopdf}
\usepackage{wrapfig}
\usepackage{natbib}
\bibpunct{(}{)}{;}{a}{}{,}

\usepackage{color}
\definecolor{darkgreen}{RGB}{0,142,128}
\definecolor{darkblue}{RGB}{0,100,170}

\begin{document}
%\title{Recipe for a good potential extrapolation with a spectropolarimetric map}
\title{From Solar to Stellar Corona: the role of wind, rotation and magnetism}

\author{Victor R\'eville$^1$,
        Allan Sacha Brun$^1$,
        Antoine Strugarek$^{2,1}$,
        Sean P. Matt$^3$,
        	J\'er\^ome Bouvier$^4$,
	Colin P. Folsom$^4$,
	Pascal Petit$^5$}

\affil{$^1$Laboratoire AIM, DSM/IRFU/SAp, CEA Saclay, 91191 Gif-sur-Yvette Cedex, France; victor.reville@cea.fr, sacha.brun@cea.fr\\
$^2$ D\'epartement de physique, Universit\'e de Montr\'eal, C.P. 6128 Succ. Centre-Ville, Montr\'eal, QC H3C-3J7, Canada; strugarek@astro.umontreal.ca\\
$^3$ Department of Physics and Astronomy, University of Exeter, Stocker Road, Exeter EX4 4SB, UK; s.matt@exeter.ac.uk\\
$^4$ IPAG, Universit\'e Joseph Fourier, B.P.53 F-38041 Grenoble Cedex 9-France, jerome.bouvier@obs.ujf-grenoble.fr, colin.folsom@obs.ujf-grenoble.fr\\
$^5$ IRAP, CNRS - Universit\'e de Toulouse, 14 avenue Edouard Belin 31400 Toulouse - France, pascal.petit@irap.omp.eu\\}

\begin{abstract}
Observations of surface magnetic fields are now within reach for many stellar types thanks to the development of Zeeman-Doppler Imaging. These observations are extremely useful for constraining rotational evolution models of stars, as well as for characterizing the generation of magnetic field. We recently demonstrated that the impact of coronal magnetic field topology on the rotational braking of a star can be parametrized with a scalar parameter: the open magnetic flux. However, without running costly numerical simulations of the stellar wind, reconstructing the coronal structure of the large scale magnetic field is not trivial. An alternative -broadly used in solar physics- is to extrapolate the surface magnetic field assuming a potential field in the corona, to describe the opening of the field lines by the magnetized wind. This technique relies on the definition of a so-called source surface radius, which is often fixed to the canonical value of $2.5R_{\odot}$. However this value likely varies from star to star. To resolve this issue, we use our extended set of 2.5D wind simulations published in 2015, to provide a criteria for the opening of field lines as well as a simple tool to assess the source surface radius and the open magnetic flux. This allows us to derive the magnetic torque applied to the star by the wind from any spectropolarimetric observation. We conclude by discussing some estimations of spin-down time scales made using our technique, and compare them to observational requirements.
\end{abstract}

%\keywords{Stuff, Stars: stuff}

\section{Introduction} 
\label{sec_intro}

The magnetic fields of most stars are created by convective motions and large scale flows in their envelopes, the source of a dynamo effect. Stellar parameters such as rotational period, mass and age influence this magnetic activity. These internal processes are difficult to probe, and must be investigated through indirect techniques such as asteroseismology. However the surface manifestation of this magnetic field largely shapes the structure of stellar coronae. Studying the Sun's corona greatly improved understanding of stellar atmospheres. Magnetic processes are thought to heat the corona up to several million Kelvin, hence driving a magnetized outflow into interplanetary space \citep{Parker1958}. As a consequence, the Sun has an expanding atmosphere, the solar wind. Within the corona, the competition between the expanding outflow and magnetic forces leads to open field regions and closed magnetic loops or streamers. To reproduce this structure, models such as the potential field source surface model (PFSS) \citep{Schatten1969} have been developed. This model assumes a current free magnetic field up to a source surface beyond which all the field lines are opened by the wind. Solar wind properties observed by spacecraft at 1 AU, have lead to the development of empirical models using the PFSS, such as the Wang-Sheeley-Arge (WSA) model \citep{WangSheeley1995}, while efforts towards more self-consistent solar wind models have been undertaken \citep[see the review of][]{HansteenVelli2012}. The WSA model uses a value of 2.5 $R_{\odot}$ as the radius of a spherical source surface ($r_{ss}$), which has been chosen to best match the polarity of the interplanetary magnetic field observed at 1 AU \citep{AltschulerNewkirk1969,Hoeksema1983} and has been extensively used ever since \citep{WangSheeley1994,SchrijverDeRosa2003,DeRosa2012}. However it has been proposed that this value should change during the Solar cycle \citep{Lee2011} due to variations of the solar magnetic activity.

Our knowledge of the Sun gives us precious insights into the coronae of other stars. Zeeman-Doppler Imaging \citep{DonatiBrown1997,Donati2006} uses the polarization of light in a line profile, produced by Zeeman splitting, to study a stellar magnetic field.  The rotationally modulated variability of that line profile, provides information about the strength and geometry of the magnetic field. The deduced surface magnetic field can in turn be used as an input for coronal models, from which integrated parameters can be estimated. In particular, observationally calibrated mass loss rates and open magnetic fluxes are key for better understanding the rotational evolution of stars. For instance, \citet{Reville2015} showed that the open flux is the relevant parameter to account for the magnetic topology in the braking induced by stellar winds \citep{Schatzman1962,WeberDavis1967,Kawaler1988,Matt2012}. A full magnetohydrodynamic (MHD) simulation is able to recover the coronal structure of such stars, however the much simpler PFSS model is likely to reproduce most of the large scale coronal magnetic field properties \citep{Riley2006}. For this latter technique, the relevant parameter -in addition to the surface field- is the source surface radius. This model has been applied to ZDI targets, with source surface radii set to different values, sometimes thanks to prominences observations, sometimes in a more arbitrary fashion \citep{Jardine2002,Jardine2013}. Given the differences of the stellar parameters: coronal temperature, rotation rate, magnetic field strength and topology, is the fiducial solar value of 2.5 $R_{\odot}$ (or $2.5 R_*$ in a stellar context) a good choice ? How can one \textit{a priori} set a reasonable value for the source surface radius ?

We investigate this by comparing the 60 MHD simulations performed in \citet{Reville2015} to potential extrapolations and estimate the optimal source surface radii matching the open flux of the stellar coronae. This is described in Section \ref{Comparison}. In Section \ref{Prediction}, we propose a general method to estimate \textit{a priori} the optimal $r_{ss}$ from stellar parameters, without running any simulations. We find that for fast rotators, magneto-centrifugal acceleration is key for assessing a correct value. We use a procedure based on \citet{Sakurai1985} (detailed in appendix \ref{AppA}) to obtain the right velocity profile, taking into account rotation and magnetic field. In Section \ref{Ccl} we discuss some applications of our method on spin-down time scale for young stars, and summarize our conclusions.

\section{Comparison of self-consistent MHD Wind simulations and potential magnetic field extrapolation}
\label{Comparison}

\subsection{Wind simulations with the MHD code PLUTO}

For more than two decades MHD simulations have been used to study the properties of stellar winds \citep{WashShib1993,KG1999,MB2004,MP2008}. Computing power has allowed for the inclusion of complex magnetic field topologies in those simulations in two or three dimensions \citep{Cohen2011,Vidotto2014,Strugarek2014sf2a}. In \citet{Reville2015}, we presented a set of 60 2.5D ideal MHD simulations to study the impact of the magnetic field topology on the stellar wind braking. We used the set of parameters of \citet{Matt2012} but extended it to more complex topologies than the dipole, such as the quadrupole and the octupole, as well as combinations of multipoles. From this study, we generalized the law giving the magnetic torque created by a wind on a solar-like star:

\begin{equation}
\tau_w = \dot{M}_w^{1-2m} \Omega_* R_*^{2-4m} K_3^2 \left( \frac{\Phi_{open}^2}{v_{esc}(1+f^2/K_4^2)^{1/2}}\right)^{2m},
\label{form}
\end{equation}
where $\dot{M}_w$ is the mass loss due to the wind, $R_*$, $\Omega_*$, and $f \equiv \Omega_* R_*^{3/2}(GM_*)^{-1/2}$ are the stellar radius, rotation rate and break-up ratio, respectively. $K_3$, $K_4$ and $m$ are the fitted parameters for the braking law. $\Phi_{open}$ is the value of the unsigned magnetic flux if the integration surface $S_r$ contains all closed magnetic loops (and is therefore a constant):

\begin{equation}
\Phi(r) = \int_{S_r} |\vec{B} \cdot d\vec{S}|
\end{equation}

This formulation and the associated coefficients have been derived using the grid of 60 numerical simulations of \citet{Reville2015}, computed with the PLUTO code \citep{Mignone2007}. All details about the numerical aspects of the study are given in \citet{Reville2015}, especially the necessary boundary conditions to properly compute the torque created by the wind. However, the formulation (\ref{form}) becomes useful to compute the torque of a given star only if $\Phi_{open}$ is known from stellar parameters. Running MHD simulations gives this value and a measure of the angular momentum loss. However, the goals here is to provide a simple method to compute this quantity without having to run time consuming simulations.

A general method that has been used widely in the solar physics community is the potential field extrapolation, which recovers the structure of the magnetic field up to a source surface radius, and assumes the wind has made the field completely radial beyond this point.

\subsection{Potential extrapolation with a source surface}

Introduced by \citet{Schatten1969}, the Potential Field Source Surface (PFSS) model is able to extrapolate the whole spatial structure of a magnetic field in a corona given the surface magnetic field. This model assumes that the magnetic field is current free in a shellular volume delimited by the stellar surface (of radius $r=r_*$) and a source surface of radius $r_{ss}$. Beyond this surface, the model mimics the effect of the wind, which opens field lines, by setting the magnetic field to be purely radial. Thus, in the region $r_* \leq r \leq r_{ss}$, we have:

\begin{equation}
\nabla \times \mathbf{B} = 0,
\end{equation}
hence there exists a scalar field $\Phi$, a potential of the magnetic field which satisfies:

\begin{equation}
- \nabla \Phi = \mathbf{B}.
\end{equation}

Since $\nabla \cdot \mathbf{B}=0$, $\Phi$ is a solution of the Laplace's equation and we can write:

\begin{equation}
\Delta \Phi = 0.
\end{equation}

Two conditions fix the value of the potential and thus the magnetic field in the whole domain. First the potential should match the observed field at the surface of the star:

\begin{equation}
\frac{\partial \Phi}{\partial r} |_{r=r_*} = - B_r (r_*, \theta, \phi).
\end{equation}

Then the potential must not depend on $(\theta,\phi)$ at $r=r_{ss}$ since the field is purely radial from this point. This implies

\begin{equation}
\Phi(r=r_{ss})=0.
\end{equation}

\begin{figure*}
\begin{tabular}{ccc}
\includegraphics[scale=0.95]{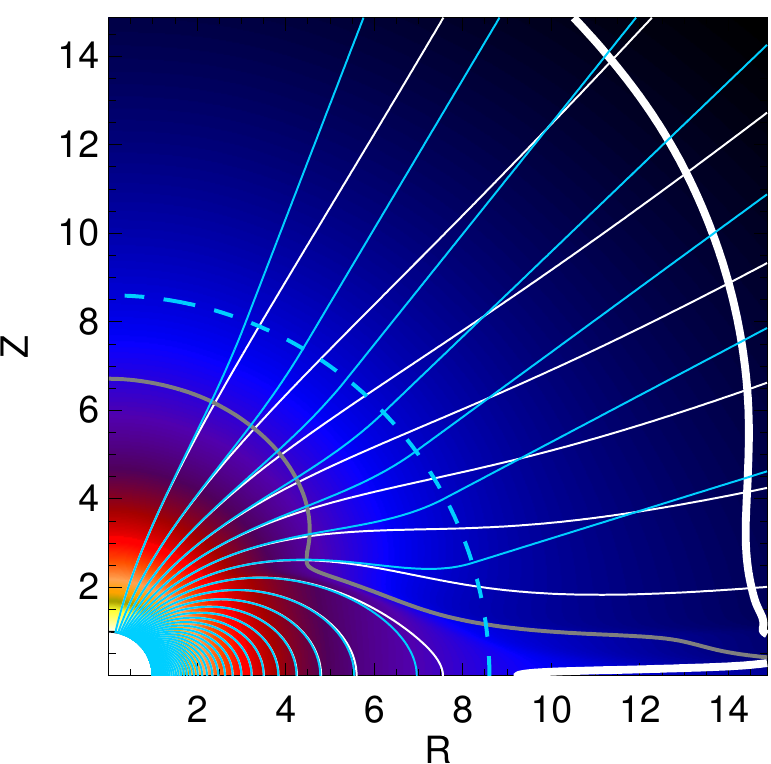} &$\quad \quad \quad$& \includegraphics[scale=0.95]{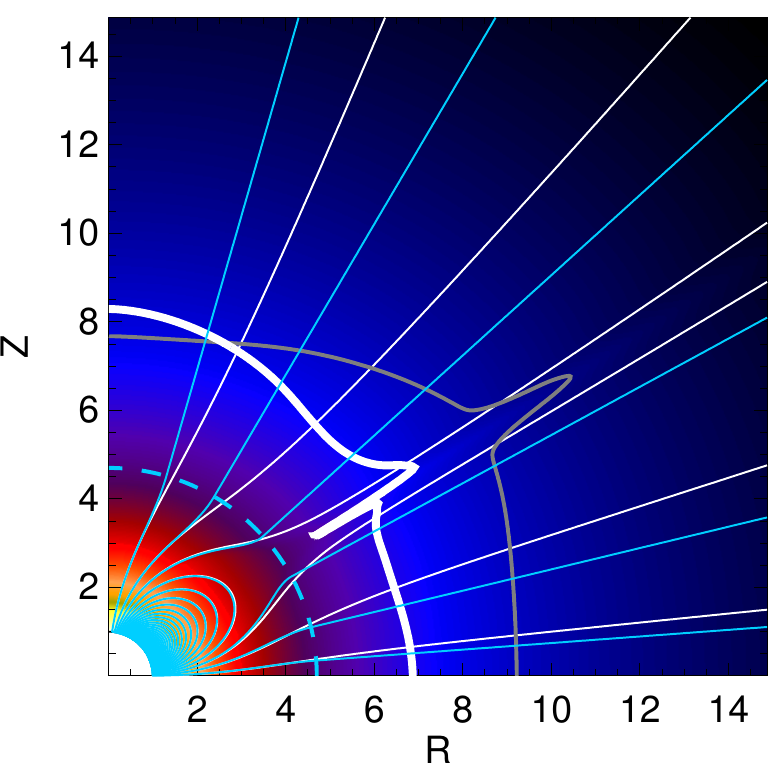}\\
\end{tabular}
\caption{Comparison of the magnetic field lines obtained by the simulations in white and by the potential extrapolation with the optimal source surface (cyan lines), in the dipolar case (left panel) and the quadrupolar case (right panel). The magnetic loops are well reproduced, but the field lines obtained by the simulations are not purely radial right beyond $r_{ss,opt}$ whereas they are -by construction- with the potential extrapolation. The cyan dashed line represent the optimal spherical source surface. Color background is the logarithm of the density. Grey and white lines are the sonic and the Alfv\'en surfaces respectively, note that the Alfv\'en surface is always beyond the source surface.}
\label{potex}
\end{figure*}

We used the derivation given in \citet{SchrijverDeRosa2003} to implement our reconstruction. In this model the magnetic field only depends on the stellar surface field and on the value of the source surface radius $r_{ss}$. Beyond this point the wind has been able to open all the field lines, which assumes that the thermal and turbulent pressure are enough to counter the tension of the coronal magnetic field. We will see in the next section how to predict a value for $r_{ss}$.

\subsection{Criteria for an optimal source surface}

In order to assess what is the best value for $r_{ss}$ and try to predict it, we need to have a reference value. For this we define an ``optimal" source surface that best describes the results of our 60 simulations. Since the quantity we are interested in is the open flux, we call the optimal source surface radius the zero of the function:

\begin{equation}
F(r_{ss})= \Phi_{open}(r_{ss}) - \Phi_{open,sim}
\label{ObjF}
\end{equation}

\begin{deluxetable}{lllccc}
  \tablecaption{Table of parameters and computed optimal source surface}
  \tablecolumns{6}
  \tabletypesize{\scriptsize}
  \tablehead{
    \colhead{Case}&
    \colhead{$v_{A}/v_{esc}$} &
    \colhead{$f$} & 
    \colhead{} & 
    \colhead{$r_{ss,opt}$} & 
    \colhead{}
  }
  \startdata
  &&& Dip. & Quad. & Oct. \\
  \hline
  \vspace{0.1cm}\\
  1	&0.0753 	&9.95e-5	&5.2		&3.3		&2.7\\
  2	&0.301	&9.95e-5	&8.6		&4.7		&3.5\\
  3	&1.51	&9.95e-5	&17.8	&7.0		&4.8\\
  3+	&2.00	&9.95e-5	&21.4	&7.7		&5.1\\
  5	&0.0753 	&9.95e-4	&5.2		&3.3		&2.7\\
  6	&0.301	&9.95e-4	&8.6		&4.7		&3.5\\
  7	&1.51	&9.95e-4	&17.8	&7.0		&4.8\\
  8 	&0.0753	&3.93e-3	&5.2		&3.3		&2.7\\
  10 	&0.301	&3.93e-3	&8.5		&4.7		&3.5\\
  13	&1.51	&3.93e-3	&17.4	&7.0		&4.8\\
  23	&0.0753	&4.03e-2	&4.3		&3.2		&2.6\\
  24	&0.301	&4.03e-2	&6.4		&4.3		&3.3\\
  25	&1.51	&4.03e-2	&9.7		&6.3		&4.6\\
  31	&0.301	&5.94e-2	&5.6		&4.1		&3.2\\
  37	&0.301	&9.86e-2	&4.3		&3.4		&2.9\\
  45	&0.301	&1.97e-1	&3.0		&2.7		&2.4\\
  47	&1.51	&1.97e-1	&4.6		&3.6		&3.2\\
  48	&0.753	&4.03e-1	&2.3		&2.2		&2.1\\
  49	&1.51	&4.03e-1	&3.0		&2.8		&2.3\\
  50	&3.01	&4.03e-1	&3.7		&3.0		&2.7\\
  \enddata
  \tablecomments{We report the optimal source surface found by comparing potential extrapolation and simulation for all the cases of \citet{Reville2015}. The values vary with the stellar parameters: the rotation rate, the magnetic field strength and the magnetic field topology.}
  \label{Table1}
\end{deluxetable}

There is a unique solution since the open flux obtained through the potential extrapolation is a decreasing function of $r_{ss}$ that starts at the surface flux value at $r_{ss}=r_*$ and tends to zero as $r_{ss}$ tends to infinity. We have been through all our simulations, and Table \ref{Table1} gives the corresponding optimal source surface radius as a function of the stellar parameters. The parameter $v_A/v_{esc}=B_*/\sqrt{4 \pi \rho_*} /\sqrt{2 GM_*/r_*}$ is the Alfv\'en speed on the surface at the equator over the escape velocity, which characterizes the magnetic field strength, while the break-up ratio $f$ characterizes the rotation rate. The other parameters used for the simulations have been kept fixed, the sound speed at the base of the corona over the escape velocity has the value $c_s/v_{esc}=0.222$ and $\gamma=1.05$. The optimal $r_{ss}$ can be easily found thanks to a bisection or a numerical Newton-Raphson method. 

Figure \ref{potex} shows a comparison of the magnetic field obtained with the simulation (white lines) and with the potential extrapolation (cyan lines) using the optimal $r_{ss}$ for dipolar and quadrupolar topologies. It can be seen that the closed magnetic loops are well reproduced. The location of the optimal source surface fairly well matches the size of the largest closed coronal loop of the simulation even though this is not the criteria we chose to define it. 

A difference can be noticed when looking at open field lines, while the potential extrapolation of the magnetic field is purely radial beyond $r_{ss}$, the solution of the simulation has field lines that bend more gradually as the wind expands. The magnetic field lines can also be collimated towards the rotation axis for high rotation \citep{Sakurai1985,Ferreira2013,Reville2015}, whereas this is not taken into account with a potential extrapolation. The potential source surface model becomes inaccurate around the optimal source surface and the deviation from the wind solution grows with larger distances even though both solutions become radial.

It seems that globally, the potential extrapolation overestimates the flux tube expansion, due to its inherent constraint to be radial beyond the source surface. At the pole however, the flux tube expansion is underestimated, at least for slow rotation. This could have consequences on solar wind models which derive the solar wind terminal speed using the expansion factor \citep{WangSheeley1990,WangSheeley1991} or the topology of the coronal field \citep[][and references therein]{Titov2012} using potential extrapolations \citep[see][]{Cohen2015SoPh}. However, as far as the open flux is concerned, it is always possible to find the optimal source surface radius that matches the simulation value.

We first notice in Table \ref{Table1} that the optimal source surface varies with the stellar parameters (magnetic field strength, topology and rotation rate). Hence the fiducial value for the Sun, $2.5 R_{\odot}$, often chosen in the literature does not correctly predict the open flux for rapidly rotating stars with strong magnetic fields. As suggested in \citet{Lee2011}, different values could be used for the Sun, whose topology varies during one cycle (see \citet{PintoBrun2011} for a detailed study of the impact of the 11 year solar cycle on the wind properties).

The value of the optimal source surface radius shows three clear trends, which are similar to the variations of the average Alfv\'en radii computed in \citet{Reville2015}. First, the optimal source surface radius grows with the magnetic field strength. Second, it decreases with higher order topology. As we will see in section \ref{Prediction}, what determines the opening of the field lines is a competition between the magnetic forces and the thermal and ram pressure of the gas. The magnetic forces that confine the gas are proportional to the surface strength and follow the radial decay imposed by the topology. Third, we can see that rotation plays a role. This is due the magneto-centrifugal acceleration \citep{WeberDavis1967,Sakurai1985,Mestel1968,Ustyugova1999}. Specifically, the optimal value of $r_{ss}$ decreases with higher rotation for a given $v_A/v_{esc}$. The ram pressure can be significantly raised by rotation \citep[see][]{Reville2015} and the wind is able to open field lines closer to the star.

\section{Prediction of the open flux and consequences on the source surface location}
\label{Prediction}

In the previous section, we have seen that the value of $r_{ss}$ can be set such as to recover the correct amount of open flux. In this section we propose a method to find an estimate of this optimal source surface radius from stellar parameters. To do so we assume that a pressure balance is established between the flow and the magnetic field at the source surface and test this criteria with two simple wind models.

\subsection{Polytropic acceleration}
In \citet{Reville2015} we made the assumption that for a solar like star, the wind is driven by the pressure gradient of an approximately $10^6$ K corona. We model this through a polytropic equation of state, mimicking a heating with a value of $\gamma=1.05$ \citep{WashShib1993,KG1999,MP2008,Matt2012}. The solution for a one dimensional, hydrodynamic (without magnetic field) polytropic wind can be computed with a Newton-Raphson method, and this is how we initialize our simulations. Complications occur when a magnetic field is introduced, especially with a complex topology. Semi-analytical methods have only been able to solve the problem with split-monopole topologies: \citet{WeberDavis1967} did so with one dimension and later \citet{Sakurai1985} extended this result into two dimensions. 

In our case, we want to give an estimate for the optimal source surface radius without having to run a MHD simulation. As we discussed earlier, in a wind solution the field lines open due to the ram and thermal pressure of the gas. The source surface was originally described \citep{Schatten1969} as the radius where the transverse magnetic energy density becomes lower than the thermal energy density. From our one dimensional polytropic profile of the speed, density and pressure, we can assess the properties of the gas, and compare them with a ``no-outflow" configuration of the magnetic field where none of the field lines are open (which is equivalent to moving $r_{ss}$ towards infinity). We then look for an equilibrium between the ram and thermal pressures and the magnetic pressure\footnote{The magnetic pressure can be written as a tensor whose maximum amplitude opposed to the gradient of the magnetic field is $B^2/2\mu_0$ \citep{PlasmaPhys}.}.

\begin{figure}
\includegraphics[scale=0.5]{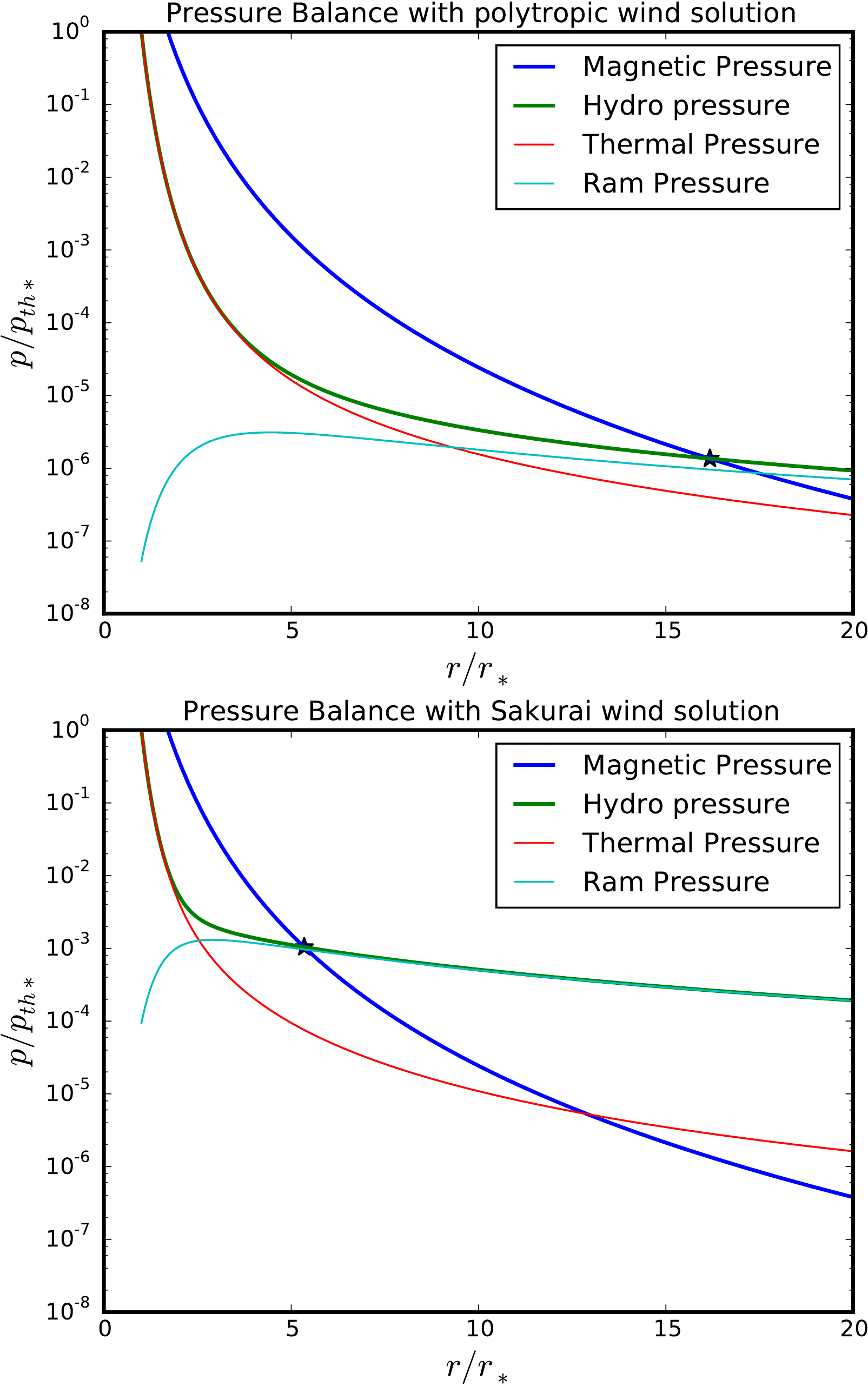}
\caption{Differences in the pressure comparison producing the estimation of the source surface radius (where the magnetic pressure crosses the hydro pressure) for the same $v_A/v_{esc}=1.51$ and a dipolar topology. In the upper panel, the magneto-centrifugal effect is not taken into account, and the velocity profile is a solution of a polytropic wind. In the lower panel we used Sakurai's formalism to derive the velocity profile in the equatorial plane with the rotation rate of case 47. We see that the acceleration is radically different, hence the estimation of the source surface radius is much smaller when magneto-centrifugal effect is included (for fast rotators).}
\label{prbal}
\end{figure}

The process is described as follows. The equation:

\begin{equation}
P_{hydro} \equiv P_{th}+P_{ram} = P_{mag},
\label{prbal1}
\end{equation}
which is equivalent to:

\begin{equation}
p+\rho v^2 = \frac{B^2}{2 \mu_{0}},
\label{prbal2}
\end{equation}

describes a surface in a 3D space. Our estimate is then simply the average spherical radius of these surface points. We will refer to it as the estimated source surface radius $r_{ss,est}$ as opposed to the optimal source surface radius $r_{ss,opt}$ computed from our simulations. This search for the pressure balance is shown in Figure \ref{prbal} in the upper panel. We can see that the thermal pressure is dominant close to the star, but the ram pressure takes over after a few stellar radii. The hydrodynamic pressure then crosses the magnetic pressure, hence defining the source surface radius estimate (although this is only a 1D profile). As a consequence, the acceleration of the wind is key to determine this value, but this model does not yet take into account the magneto-centrifugal acceleration due the magnetic field lines anchored to the rotating star. 

\subsection{Magneto-centrifugal acceleration}
\label{Rotation}

The magneto-centrifugal effect is a simple consequence of the existence of a star's coronal magnetic field. \citet{Schatzman1962} imagined that a magnetized wind could carry angular momentum, and thus could be responsible for main sequence stars braking. This concept has been further quantified by \citet{WeberDavis1967} who described the magnetic field as a lever-arm acting on the star's rotation. Anchored to the rotating star, the magnetic field lines drag the gas in their rotation so that the gas feels a centrifugal force and is accelerated. This acceleration can be equivalent to, or even higher than the one due to the pressure gradient in thermal winds, leading to the slow and fast magnetic rotators theory \citep[see][]{BelcherMacGregor1976,LC1999}.

\begin{figure}
\includegraphics[scale=0.47]{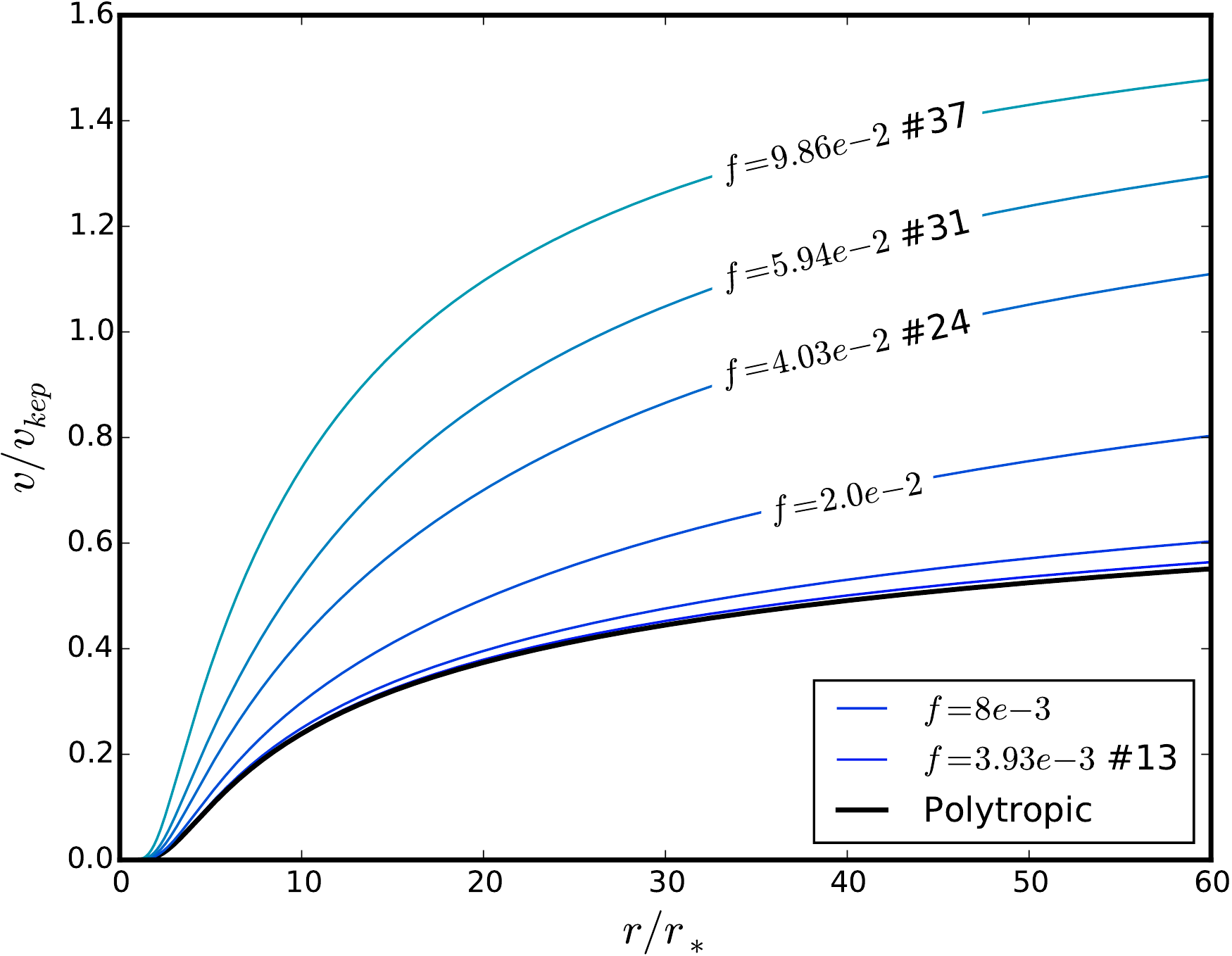}
\caption{Comparison of the wind speed profiles in various cases of our study with a fixed magnetic field strength. The polytropic solution is the lowest while Sakurai Solution grows with higher rotation rates. Sakurai and polytropic solutions begins to differ with case 13 at $f=0.00393$.}
\label{Sak2Poly}
\end{figure}

\citet{WeberDavis1967} used a semi-analytical method to compute their solution, which yields the toroidal and the poloidal velocity and magnetic fields near the equatorial plane for a purely radial magnetic field. However, to take into account the magneto-centrifugal effect we chose to implement the formalism used in \citet{Sakurai1985}. This formalism solves the same problem but with an improved methodology. We considered only the 1D poloidal profile in this work. Details of the implementation of this method are given in appendix \ref{AppA} for interested readers. In \citet{Reville2015}, we showed that rotational effects begin to be important for wind acceleration when $f \geq 0.01$. To understand the strong effect of magneto-centrifugal acceleration on the velocity profile, we plot different wind speed solution profiles for different rotation rates obtained through Sakurai's method in Figure \ref{Sak2Poly}. As expected, the shown velocity profiles tend to the polytropic solution as rotation rate decreases. We can see that the solution begins to differ with the polytropic profile for case 13 ($f=0.00393$). For case 37, the velocity amplitude is almost four times the polytropic one at 10 $r_*$. As a consequence, we expect the ram pressure ($p_{ram}=\rho v^2$) to be strongly modified by the magneto-centrifugal effect.

Coming back to Figure \ref{prbal}, we see how the pressure balance is modified when the magneto-centrifugal effect is included (bottom panel). We take the case of a fast magnetic rotator ($f=0.197,\quad v_A/v_{esc}=1.51$) and the one dimensional profiles are in the equatorial plane where the magneto-centrifugal effect is maximum. The ram pressure is very significantly raised in the lower panel and the $r_{ss,est}$ derived from this pressure comparison is thus much closer to the star. The thermal pressure is raised as well, due to more energy injected in the system, although this appears to be less significant for our pressure balance. The magneto-centrifugal acceleration is also larger for a higher magnetic field strength, at a given rotation rate. 

\subsection{Results}

\begin{figure}
\includegraphics[scale=0.6]{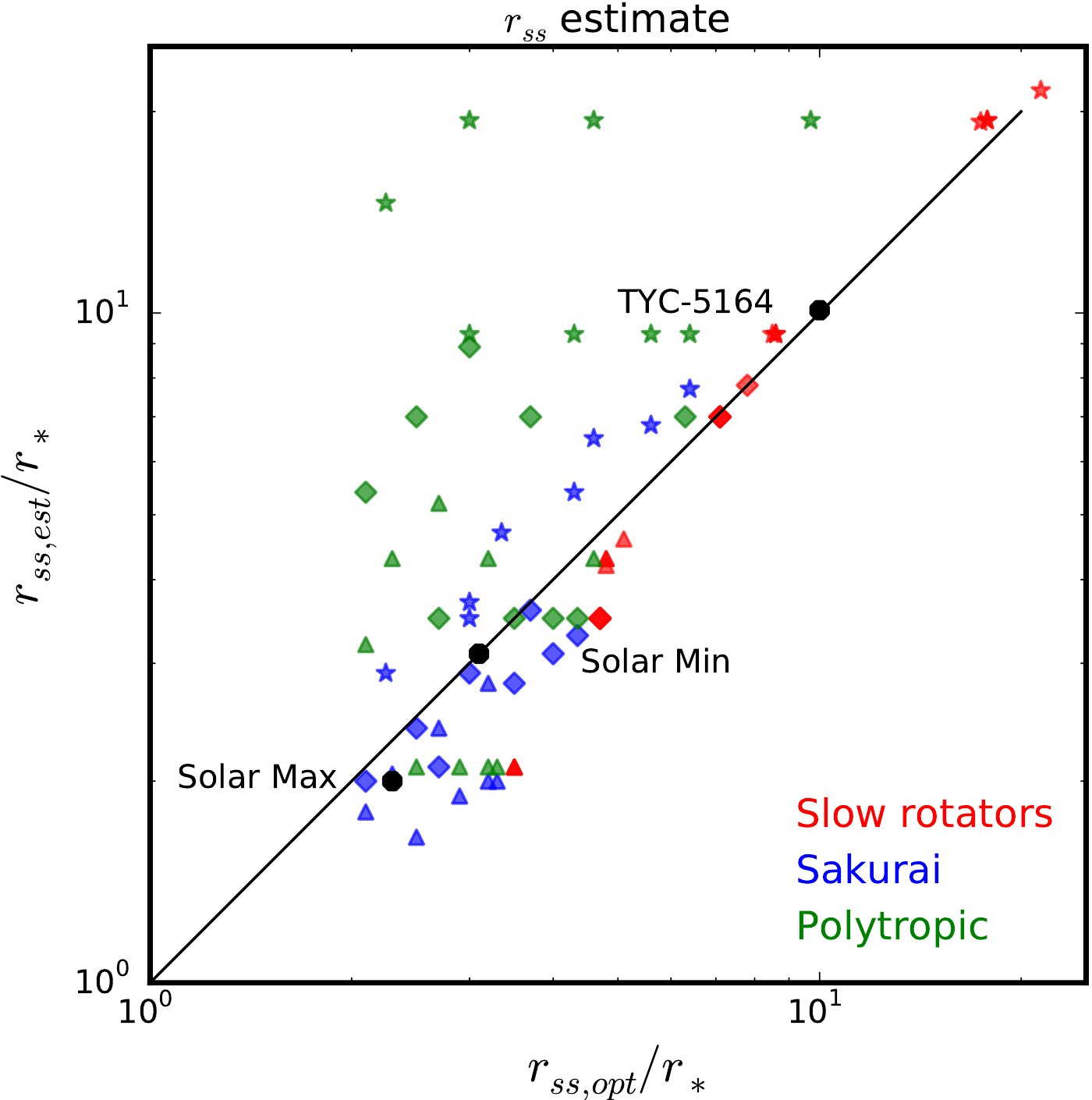}
\includegraphics[scale=0.6]{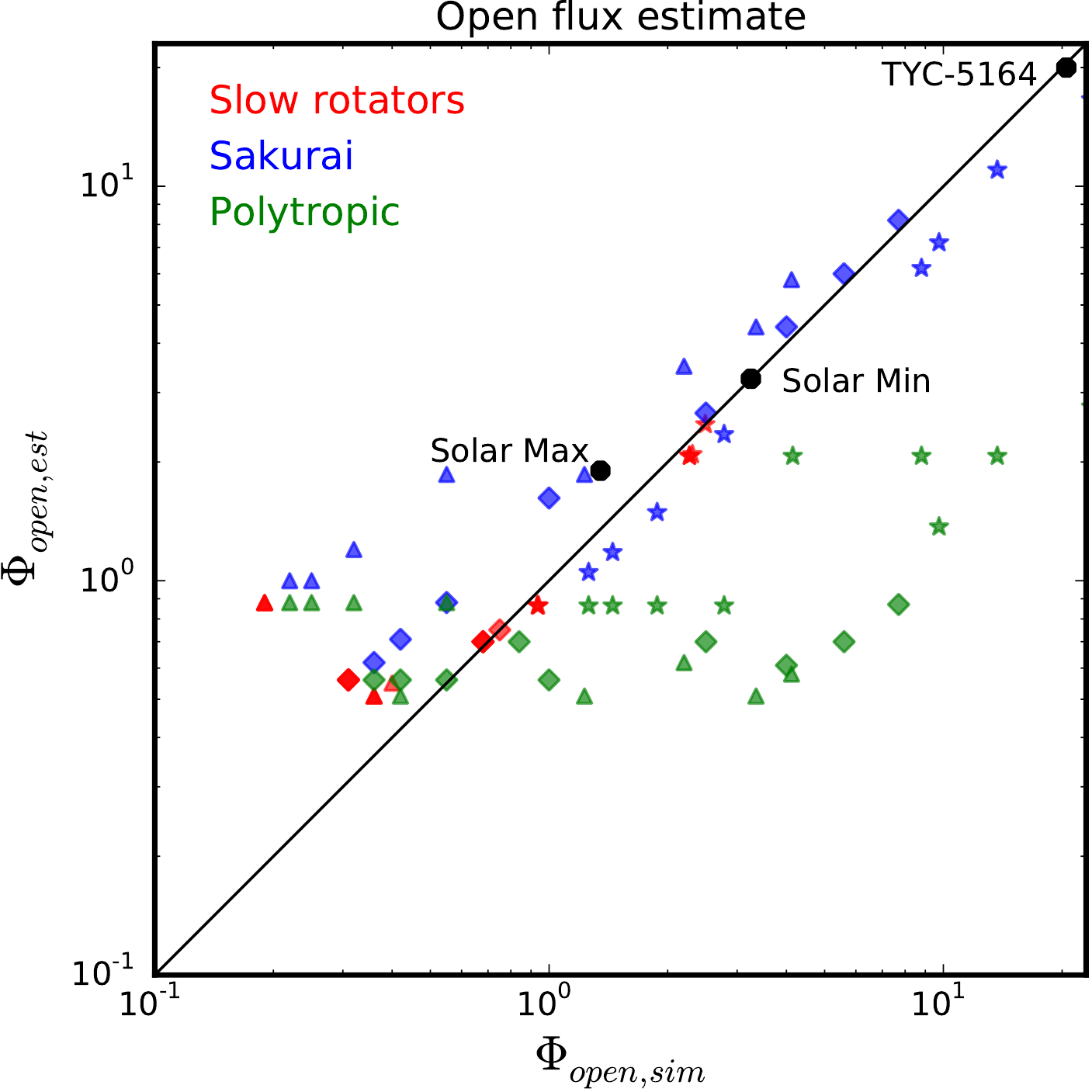}
\caption{Comparison of the estimate of $r_{ss}$ (upper panel) and the corresponding open flux (lower panel) depending on whether the magneto-centrifugal effect is taken into account. For slow rotators ($f \leq 10^{-2}$, red points), there is no difference and we notice a rather good agreement between the prediction and the optimal value. For fast rotators, we see a large mismatch if we assume that the acceleration of the wind is only given by thermal gradient through a polytropic equation of state (green points). The agreement is better if we use the magneto-centrifugal wind prescription (blue points). The symbols stand for the topology: stars for dipoles, diamonds for quadrupoles, and triangles for octupoles. Mixed topology cases are represented with black octogones.}
\label{scatter}
\end{figure}

Figure \ref{scatter} shows the estimates of $r_{ss}$ and the deduced open fluxes for our set of parameters using both the polytropic and Sakurai wind solutions, and compares them to the outputs of the numerical simulations. For both panels, red points represents slow rotators ($f \le 0.01$) and for fast rotators we distinguish between the two wind models we implemented. Green points are computed with the polytropic model while blue points include magneto-centrifugal acceleration through Sakurai's technique. The top panel compares the optimal source surface radius for a given case with the source surface radius estimate derived from formulae (\ref{prbal1})-(\ref{prbal2}). We can see that, while both radii are close at slow rotation (red symbols), for faster rotation the Sakurai wind model gives much more accurate estimates than the polytropic one. For fast rotation rates and large magnetic fields, the estimates obtained with the Sakurai wind model are three to four times more accurate than the estimates obtained by the polytropic wind model. With the Sakurai wind model, the average relative error\footnote{The relative error is defined by: $err = 2 | r_{ss,opt}-r_{ss,est} | /( r_{ss,opt}+r_{ss,est})$} of this technique is around 20\% (60 \% with the polytropic wind model).

The bottom panel compares the estimate of the open flux computed with $r_{ss,est}$. It is interesting to note that the overall shape of this plot is inverted compared to the above since the open flux computed from a potential extrapolation decreases with increasing $r_{ss}$. Once again the Sakurai model gives more accurate estimates. For large values of the open flux ($\Phi_{open,sim}  \geq 1$) that significantly raise the magnetic torque (see formulation \ref{form}), the polytropic wind model almost systematically fails to reproduce the simulation value with relative error that can reach 160\%, when we stay below 20\% with the Sakurai model. However, pure octupolar cases with weak magnetic fields are hard to catch with both models but they are unlikely to occur in realistic magnetic field configurations.

Figure \ref{scatter} show some trends with the topology. For instance, in the upper panel, dipolar points with the Sakurai wind model (blue stars) are a little above the $y=x$ line. In this case, the streamer is located at the equator where the magnetic field strength is lower and the magneto-centrifugal effect is maximum. Hence the pressure balance at the equator gives slightly better estimates. Octupolar cases with both models (triangles) are usually below the $y=x$ line. This leads to overestimation of the open flux. It is particularly true for weak magnetic fields. It seems that there is a saturation of $r_{ss,opt}$ around $2R_*$, \textit{i.e.} the value does not go below this, while the pressure equilibrium can occur down to $1R_*$ or even yield no results if the hydro pressure is always higher than the magnetic pressure. This occurs in our study, for all the cases with $v_A/v_{esc}=0.0753$. Those points are thus not represented in Figure \ref{scatter}. Hence to use this methodology, it is necessary to ensure that the magnetic pressure is higher than the thermal pressure at the base of the corona\footnote{This condition, equivalent to $\beta < 1$ at the base of the corona, is true for the Sun where $\beta \approx 0.1$. For faster rotators with higher magnetic fields, following the prescription of \citet{HolzwarthJardine2007}, $\beta=p_{th,*}/(B_*^2/2/\mu_0) \propto (\Omega_*/\Omega_{\odot}) ^{(0.6+0.5)-2\times1.2} = (\Omega_*/\Omega_{\odot})^{-1.3}$. Hence the condition is likely to be always true for the solar rotation rate and above.}. Also, we propose not to go below a saturation value of $2R_*$ for $r_{ss,est}$ for all cases. Doing so systematically improve the estimation of $r_{ss}$ and $\Phi_{open}$.

Nonetheless, for complex magnetic fields, the criteria derived in equations (\ref{prbal1})-(\ref{prbal2}), and illustrated in Figure \ref{prbal} and \ref{scatter}, works well. We added three realistic cases that represents the Sun at its minimum of activity and at its maximum of cycle 22, and the young star TYC-5164-567-1. The magnetic field spherical harmonics coefficients for the Sun are taken from \citet{DeRosa2012} and were measured at the Wilcox Solar Observatory. For the Sun we change the value of $c_s/v_{esc}$ to $0.26$ and we consider a density at the surface of $\rho_*=1.67 \times 10^{-16}$ g/cm$^3$. This value is calibrated such that the velocity at 1 AU and the mass loss rate fit observed values for $\gamma=1.05$, \textit{i.e} around $450$ km/s and $3 \times 10^{-14} $M$_{\odot}$/yr, for both wind models (since the Sun is a slow rotator). We find that the solar $r_{ss,opt}$ at minimum and maximum obtained with our wind simulations bracket the fiducial value of $2.5 R_{\odot}$ with $r_{ss,opt}=2.1$ at maximum and $r_{ss,opt}=3.1$ at minimum. Our estimate at the minimum of activity perfectly matches the optimal value, while the estimate at the maximum $r_{ss,est}=1.7$ is slightly under the saturation value of $2 R_*$, which has been found to be the minimum size of streamers in our study.

The source surface radius is larger at solar minimum because of a much stronger dipole than during maximum, which has a strong quadrupole. Interestingly, \citet{Lee2011} predicted the opposite variation of the source surface radius between minimum and maximum of activity. This latter study focused on mid-latitude coronal holes that are non-axisymmetric and small scale features that we do not account for here, to justify the variations of $r_{ss}$. This will be investigated in the near future.

TYC-5164-567-1 is a 120 Myr-old K-star, of mass $M=0.85 M_{\odot}$ and rotational period $P=4.7$ days. We set $c_s/v_{esc}=0.285$ and $\rho_*=4.86\times 10^{-16}$ g/cm$^{3}$, which is consistent with the prescription of the coronal temperature evolution with the rotation rate given by \citet{HolzwarthJardine2007} considering the value of $\gamma=1.05$ we use. The spherical harmonics coefficients for the surface magnetic field have been obtained by ZDI, using observations from the spectropolarimeter ESPaDOnS (Echelle Spectropolarimetric Device for the Observations of Stars) at the CFHT (Canada France Hawaii Telescope) \citep{Folsom2015}. Using our method and comparing it to a MHD simulation, we find that $r_{ss,opt} = 9.8R_*$ and  $r_{ss,est}=10R_*$. This example demonstrates how inaccurate the fiducial value used for the Sun can be for other stellar targets. This large value is mainly due to a $150$ G axisymmetric dipole, which is common for such young rapidly rotating stars. The strong magnetic field also explains the large value of the open flux. 

For the three realistic cases, the correlation between the open magnetic flux and the value of $r_{ss}$ is different than for the rest of the study. Increasing the coronal temperature reduces the value of $r_{ss}$, at a given magnetic field strength due to a larger pressure gradient and more thermal acceleration. Those points demonstrates that our method is valid for different coronal temperatures, which are known to vary from stars to stars \citep{Preibisch1997,Gudel2004,HolzwarthJardine2007}.
  
\section{Discussion}
\label{Ccl}

The coronal structure of a magnetic field varies with stellar parameters. For a given coronal temperature, the magnetic streamers will grow with the intensity of the surface magnetic field. Magnetic topology also plays an important role. Hence, the PFSS model should take into account those parameters and we propose a method to do so in this paper. It might seem surprising that a one dimensional solution, which assumes a split-monopole topology, can be compared with two dimensional complex magnetic fields derived by 2.5D MHD wind simulations. This comes about because the acceleration process occurs on open field lines where, locally, the geometry of the magnetic field is close to a monopole. Hence the profile derived from the Sakurai technique is close to the one observed in simulations. This method could further be improved by considering a latitudinal dependency of the magneto-centrifugal acceleration (maximum at the equator) and the location of the streamers, particularly for 3D non-axisymmetric configurations.

For the Sun, we find that the fiducial value of $2.5 R_{\odot}$ is consistent with the optimal value we find at maximum and minimum of activity ($2.1 R_{\odot}$ and $3.1 R_{\odot}$). However for younger stars with magnetic fields that can reach the kilogauss scale, we have seen that this value can be far from the optimum. The choice of $r_{ss}$ has important consequences for the structure of the astrosphere and stellar dynamics. 

With physically based arguments, we propose here a simple way to compute the magnetic torques for any target. For the Sun, using the open flux computed from a potential extrapolation made at the $r_{ss}$ predicted by our technique, and using the mass loss from our wind solution profiles (that matches observations), we find  a spin down time scale of 17 Gyr, at the minimum of activity. At maximum the time scale goes up to 46 Gyr. Those values are in good qualitative agreement with the pioneering work of \citet{Skumanich1972} and recent studies of \citet{Matt2015} and \citet{GalletBouvier2013}, from which a spin-down time scale of $10$ Gyr or more can be expected.

For TYC-5164-567-1, we find a spin-down time scale of 400 Myr. Fits from observations of clusters suggest a value of 130 Myr \citep{Matt2015}. This estimate could be improved by taking into account non-axisymmetric modes in the potential extrapolation. More complex reconstructions are also possible. In this work we only take into account the radial component of the magnetic field, but more general methods have been proposed such as constant-$\alpha$ force free fields \citep{Berger1985}, or non-potential fields \citep{Jardine2013}. This could lead to more accurate results for realistic topologies obtained by ZDI. Moreover, our method is based on stellar parameters that are still poorly constrained for distant stars, such as the density and the temperature at the base of the corona. The heating process used in our wind solution is also fairly simple and more accurate descriptions of the physical processes, including for instance, energy inputs from Alfv\'en waves and radiative losses at the base of the corona will be implemented in the near future \citep{SchwadronMcComas2003,Suzuki2006,Velli2010}. The torque computation is very sensitive to those prescriptions. A more detailed study of the torques we get with this formulation will follow in an upcoming paper.

We believe this method is a step towards understanding the coronal properties and angular momentum loss of low-mass stars. A open source python script that will perform all the calculations given a magnetic field strength, topology, and stellar parameter $(R_*,M_*, \Omega_*, \rho_*,T_*)$, can be obtained by contacting the first author.

\section{Acknowledgements}
We thank the ANR Blanc TOUPIES SIMI5-6 020 01, the ERC STARS2 207430, and CNES via Solar Orbiter funding for their support. Antoine Strugarek is a National Postdoctoral Fellow at the Canadian Institute of Theoretical Astrophysics (CITA) and acknowledges support from the Canada's Natural Sciences and Engineering Research Council.

\bibliographystyle{yahapj}
\bibliography{biblio}

\begin{appendix}
\section{Weber and Davis Solution through Sakurai technique}
\label{AppA}

\citet{Sakurai1985} proposed a method to numerically compute the \citet{WeberDavis1967} wind model. This method solves the problem of a magnetized wind anchored to a rotating star, whose magnetic field is purely radial. The solution is solved in the $(r,\phi)$ plan at the equator, hence only the radial and azimuthal components of the magnetic and velocity fields $(V_r,V_{\phi},B_r,B_{\phi})$, and the density $\rho$ and pressure $p$ profiles are of interest. The MHD equations can then be integrated as follows:

\begin{equation}
p=K \rho^{\gamma},
\end{equation}

\begin{equation}
\rho V_r r^2 = f,
\end{equation}

\begin{equation}
B_r r^2 = \Phi,
\end{equation}

\begin{equation}
(V_{\phi}-\Omega r)B_r=V_r B_{\phi},
\end{equation}

\begin{equation}
r \left( V_{\phi}-\frac{B_r B_{\phi}}{4 \pi \rho V_r} \right) = \Omega r_A^2,
\end{equation}

\begin{equation}
\frac{V_r^2}{2}+\frac{1}{2}(V_{\phi}-\Omega r)^2+\frac{\gamma}{\gamma-1}\frac{p}{\rho}-\frac{GM}{r}-\frac{\Omega^2 r^2}{2} = E,
\label{bernoulli}
\end{equation}

where $K,f,\Phi$ and $r_A$ the Alfv\'en radius, are integration constants while $\Omega$ is the rotation rate of the star.

Manipulating those equations can lead to a formalism, developed by \citet{Sakurai1985}, which allows one to find a unique solution given the stellar parameters. This appendix focuses on a description of the numerical method and the results, over a derivation of this formalism.

Substituting all previous equations into equation \ref{bernoulli}, we obtain an equation that only depends on $r$ and $\rho$:

\begin{equation}
H(r,\rho)=E.
\end{equation}

Normalizing by quantities at the Alfv\'en point, where the wind reaches the Alfv\'en speed, we can write:

\begin{equation}
H(r,\rho)=\frac{GM}{r_A} \tilde{H}(x,y),
\end{equation}

where

\begin{equation}
x=r/r_A, \quad y=\rho/\rho_A, \quad \rho_A=4\pi f^2/\Phi^2,
\end{equation}

\begin{equation}
\tilde{H}(x,y) = \frac{\beta}{2 x^4 y^2} + \frac{\Theta}{\gamma-1} y^{\gamma-1}-\frac{1}{x}+\frac{\omega}{2} \left[\frac{(x-1/x)^2}{(y-1)^2}-x^2 \right],
\end{equation}

and

\begin{equation}
\beta = \frac{\Phi^2}{4\pi GM \rho_A r_A^3}=\left[V_{Ar}^2/\frac{GM}{r}\right]_A,
\end{equation}

\begin{equation}
\Theta=\frac{\gamma K \rho_A^{\gamma-1} r_A}{GM}=\left[C_S^2/\frac{GM}{r}\right]_A,
\end{equation}

\begin{equation}
\omega=\frac{\Omega^2 r_A^3}{GM}=\left[\Omega^2 r^2/\frac{GM}{r}\right]_A.
\end{equation}

The solution of the problem is given by the contour of $\tilde{H}=\tilde{E}=E/(GM/r_A)$, that goes through two critical points $(x_s, y_s)$ and $(x_f,y_f)$ corresponding to the slow and fast magnetosonic points. The contour gives $y(x)$ and thus $\rho(r)$. The equations satisfied by $\tilde{H}$ are:

\begin{equation}
\frac{\partial \tilde{H}}{\partial x} = \frac{\partial \tilde{H}}{\partial y} = 0, \quad \tilde{H}=\tilde{E},
\end{equation}

at two locations: $(x,y)=(x_s,y_s) \mbox{ and } (x_f,y_f)$.\\

We then obtain 6 equations and 8 unknowns $\beta, \Theta, \omega, \tilde{E}, x_s, y_s,x_f,y_f$ if we keep $\gamma$ fixed. However our unknowns are not independent, and are constrained by two more equations that depend on stellar parameters:

\begin{equation}
\tilde{E}/\omega^{1/3}=\left( \frac{\gamma}{\gamma-1}\frac{p_*}{\rho_*}-\frac{GM}{r_*}-\frac{\Omega^2 r_*^2}{2} \right)/(GM\Omega)^{2/3}= q_1
\end{equation}

\begin{equation}
\beta^{\gamma-1} \Theta/\omega^{4/3-\gamma} = \frac{\gamma p_*}{\rho_*} \left(\frac{B_{r*}^2}{4 \pi \rho_*} \right)^{\gamma-1}/\left[ \left(\frac{GM}{r_*}\right)^{2\gamma-4/3} (\Omega r_*)^{2(4/3-\gamma)} \right] = q_2
\end{equation}

The general method is thus to look for 6 parameters as a function of the two other parameters. For instance, for known $\Theta$ and $\omega$, a six dimensional Newton-Raphson can be used to find a unique solution. Let us define the function:

\begingroup
\renewcommand*{\arraystretch}{1.5}

\begin{equation}
F= \left(
\begin{array}{c}	
f_1= \tilde{H}(x_s,y_s,\beta)-\tilde{E} \\	
f_2= \tilde{H}(x_f,y_f,\beta)-\tilde{E}  \\
f_3= \partial_{x} \tilde{H} (x_s,y_s,\beta)\\
f_4= \partial_{x} \tilde{H} (x_f,y_f,\beta)\\
f_5= \partial_{y} \tilde{H} (x_s,y_s,\beta)\\	
f_6= \partial_{y} \tilde{H} (x_f,y_f,\beta)
\end{array}
\right)
\end{equation}

\endgroup
The algorithm to find the zero for this function is described as follows:

\begin{itemize}
\item Choose an initial guess $X_0$
\item $F_N=F(X_0)$
\item \textbf{while} ($F_N \geq \varepsilon$):\\
 $\quad \quad X_{N+1}=X_N-J(X_N)^{-1} F_N$\\ 
 $\quad \quad F_{N+1}=F(X_{N+1})$\\
\end{itemize}	

where $J$ is the Jacobian matrix of F taken at $X_N$. If the initial guess is close to the solution, this method is remarkably fast and efficient.

Mapping the values of $\tilde{E}$ and $\beta$ as a function of $\Theta$ and $\omega$, one can easily find the intersections of the two contour lines corresponding to the values $q_1$ and $q_2$ of the resulting arrays $\tilde{E}/\omega^{1/3}$ and $\beta^{\gamma-1} \Theta/\omega^{4/3-\gamma}$. Then, the solution is fully determined from the stellar parameters, given a choice for $\gamma$.

\begin{figure}
\center
\includegraphics[scale=0.6]{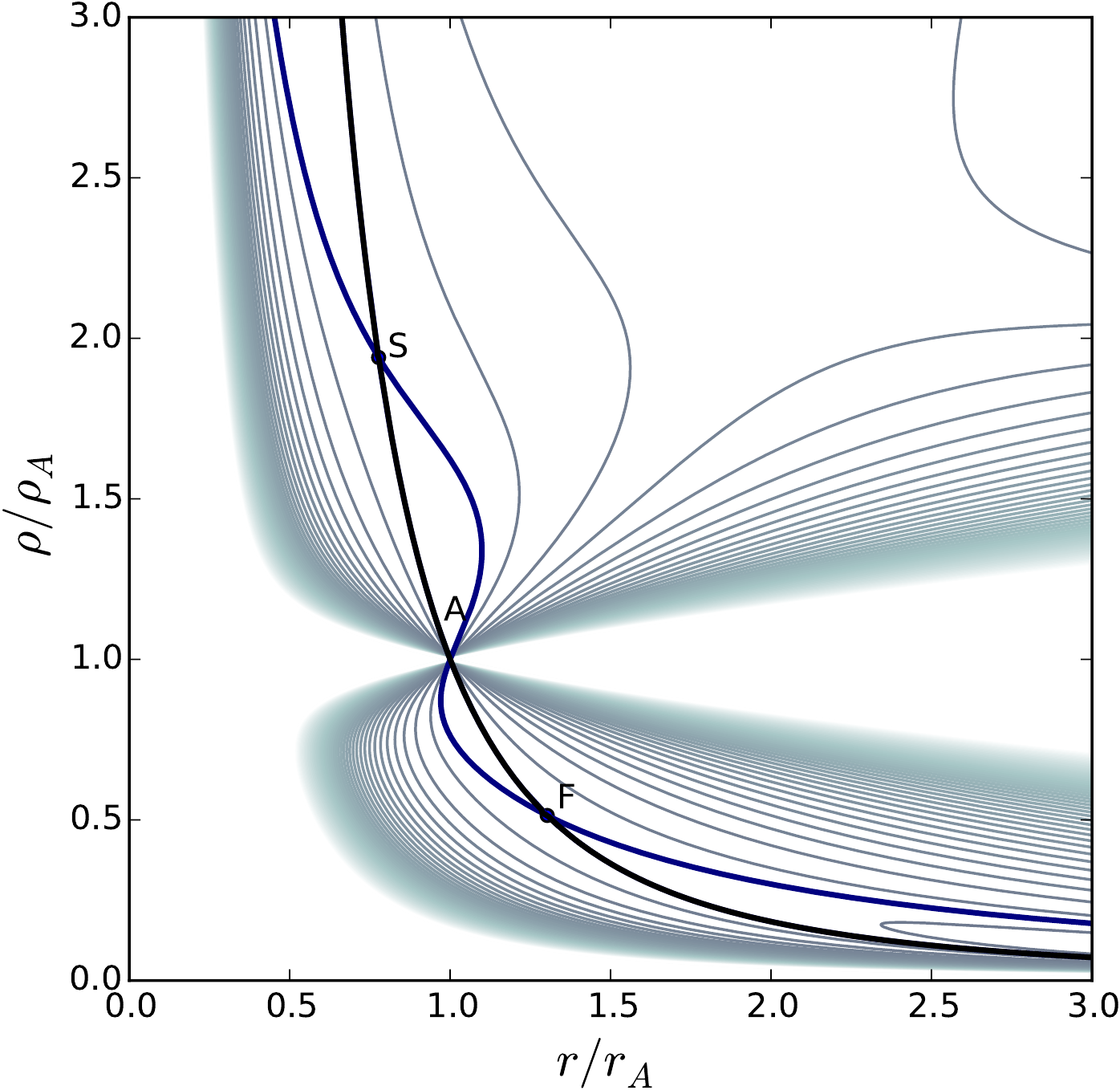}
\includegraphics[scale=0.6]{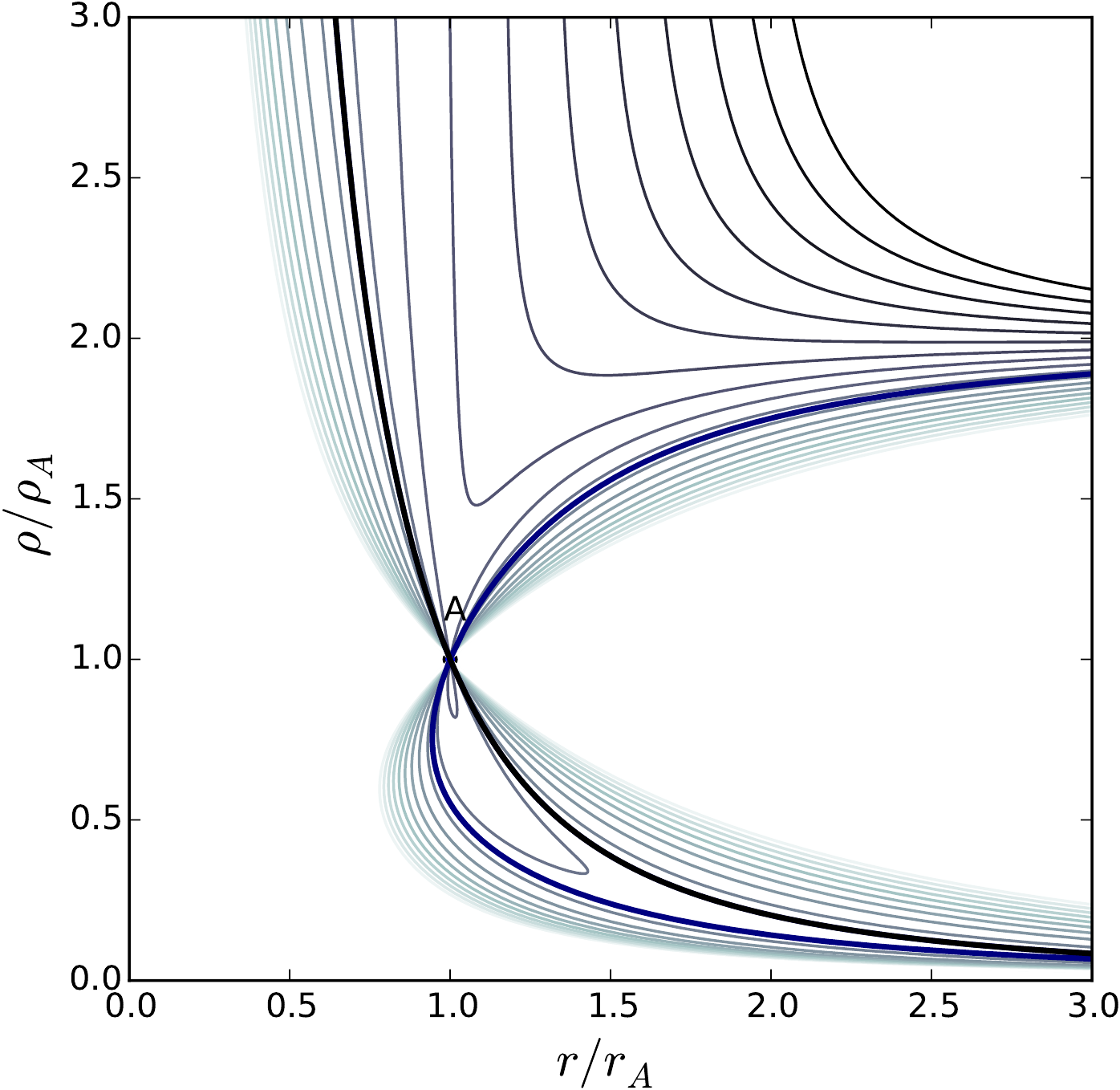}
\caption{Density profile obtained through Sakurai's formalism in the published case of $\gamma=1.2$, $\Theta=0.5$, $\Omega=0.25$ (left panel) and in the case 47 of our study corresponding to parameters $\gamma=1.05$, $\Theta=1.23$, $\Omega=550$ (right panel). In the latter case the slow and fast critical points are out of the domain $[0,3] \times [0,3]$. Grey lines are the contour lines of $\tilde{H}$, the thick lines correspond to the contour at energy $\tilde{E}$ although the black one is the only physical solution among them.}
\label{cnt}
\end{figure}

In Figure \ref{cnt} we show two density solution for different parameters, to ensure the reproducibility of our results. In the left panel is the solution published in \citet{Sakurai1985}, where the fast and slow critical points are close but clearly distinguishable from the Alfv\'en point. The full solution for this case is: 

\begin{equation}
\gamma=1.2, \quad \Theta=0.5, \quad \omega=0.25, \quad \beta=0.576, \quad \tilde{E}=1.738, \quad (x_s, y_s)=(0.777,1.940), \quad (x_f,y_f)=(1.302,0.514)
\end{equation}

In the right panel we show the density solutions for the case 47 of our study. We can see that, due to the different parameters, the structure of $\tilde{H}$ is greatly distorted so that the critical points are out of the shown domain. The full solution is given by the following parameters:

\begin{equation}
\gamma=1.05, \quad \Theta=1.235, \quad \omega=550.0,\quad \beta=167.3, \quad \tilde{E}=23.0, \quad (x_s, y_s)=(0.105,902.3), \quad (x_f,y_f)=(10.1,0.0066)
\end{equation}

\end{appendix}

\end{document}